\documentstyle[12pt]{article}
\def \beq{\begin{equation}}

\def \eeq{\end{equation}}

\begin{document}
\begin{titlepage}
\vspace{-3in}
\rightline{CERN-TH/96-169}
\rightline{EFI-96-25}
\rightline{hep-ph/9607207}
\begin{center}
{\large\bf Prominent Decay Modes of a Leptophobic $Z'$
\footnote{To be published in Physics Letters B.}} \\
\vspace{1.5cm}
{\large Jonathan L. Rosner} \\
\vspace{.5cm}
{\sl CERN, 1211-CH Geneva 23, Switzerland} \\
\vspace{.5cm}
{\sl Enrico Fermi Institute and Department of Physics}\\
{\sl University of Chicago, Chicago, IL 60637 USA}
\footnote{Permanent address.}\\
\vspace{1.5cm}
\begin{abstract}

An anomaly-free U(1) charge $Q'$ has recently been identified within the group
E$_6$ for which the familiar leptons (the left- and right-handed electron and
the left-handed neutrino) have $Q' = 0$.  It is pointed out that the $Q'$
charges of several {\it exotic} leptons within E$_6$ matter multiplets are
quite large, leading to the prediction that half of the decays of the so-called
``leptophobic'' $Z'$ bosons coupling to $Q'$ are to these exotic leptons. Other
large $Q'$ charges include those of standard up-type quarks and exotic
down-type quarks.  Substantial forward-backward asymmetries are expected in
$u \bar u \to Z' \to f \bar f$ channels when $f$ is a standard up-type quark,
an exotic down-type quark, or an exotic lepton.
\end{abstract}

\end{center}
\leftline{PACS codes: 14.70.Pw, 12.60.Cn, 12.10.Dm, 13.38.Dg}
\vfill
\leftline{CERN-TH/96-169}
\leftline{July 1996}
\end{titlepage}

It was proposed some time ago \cite{delA} that searches for new neutral gauge
bosons $Z'$, traditionally pursued at the highest energies using the reaction
$\bar p p \to Z' + \ldots \to (e^+ e^- + \ldots$ or $\mu^+ \mu^- + \ldots)$
\cite{CDFZp}, may have overlooked such bosons if they are {\it leptophobic},
i.e., if their couplings to the standard leptons are suppressed. The interest
in such $Z'$ states extends beyond recent specific motivations (see, e.g.,
Refs.~[3--12]) which seek to explain an excess branching ratio $B(Z \to b \bar
b)$ \cite{Rbb} and an excess of jets produced at high transverse momenta in
$\bar p p$ collisions \cite{pT} with respect to standard-model predictions. 

Although a leptophobic $Z'$ might appear artificial from the standpoint of
unified theories of the electroweak and strong interactions, such a state can
be constructed using the U(1) charges available within the group E$_6$
\cite{delA,BKM}.  In this note we wish to point out that this $Z'$, although
its couplings shun the traditional leptons, decays half the time to {\it
exotic} leptons which are contained within the matter multiplets of E$_6$. Such
leptons are part of the complement of fermions which are required in order that
the U(1) be anomaly-free.  We shall also note that the couplings of this state
favor up-type quarks over down-type quarks, in contrast to those of the
standard $Z$.  In contrast to Refs.~\cite{FWW}, \cite{LN}, and several others
mentioned in Ref.~\cite{BCL}, we consider for simplicity only a model with
family-independent couplings. 

Candidates for groups unifying color SU(3) with electroweak SU(2) $\times$ U(1)
include SU(5), SO(10), and E$_6$.  The SU(5) group \cite{GG} is the smallest
with which this unification can be achieved; the familiar left-handed fermions
belong to a {\bf 5}$^* + ${\bf 10}-dimensional reducible representation.  With
the addition of a right-handed neutrino, these two representations may be
combined in a single {\bf 16}-dimensional spinor of SO(10) \cite{SO}.  This, in
turn, is contained in a {\bf 27}-dimensional representation of E$_6$, which is
a group often encountered in superstring theories \cite{E6SS} but whose
possibilities for strong-electroweak unification were explored before the
superstring era \cite{E6}.  In addition to the SO(10) {\bf 16}-plet, the {\bf
27} of E$_6$ contains representations of (SO(10), SU(5)) with dimensions ({\bf
10}, {\bf 5}$^* + ${\bf 5}) and ({\bf 1}, {\bf 1}).

Extra-U(1) factors can be identified in various ways.  A maximal subgroup of
SO(10) containing SU(5) includes an additional U(1) which is conventionally
labelled U(1)$_\chi$, while a maximal subgroup of E$_6$ containing SO(10)
includes an additional U(1) called U(1)$_\psi$ \cite{chipsi,LRR,LR}.  One
particular combination of these two U(1)'s is frequently discussed in the
context of superstring theories \cite{E6SS} and is called U(1)$_\eta$.  (More
details on searches for extra $Z$'s, including $Z_\eta$, may be found in
Refs.~\cite{lore}.)  The leptophobic $Z'$ constructed in Ref.~\cite{BKM}
couples to a linear combination of U(1)$_\eta$ and the weak hypercharge
belonging to the U(1) of the standard electroweak theory. 

For our purposes it is more convenient to label the U(1) factors within E$_6$
by means of the isospins and weak hypercharges in the decomposition E$_6 \to
$SU(3)$_C \times$ SU(3)$_L \times$ SU(3)$_R \to $SU(3)$_C \times $SU(2)$_L
\times $U(1)$_L \times $SU(2)$_R \times $U(1)$_R$ \cite{E6SS}.  The {\bf
27}-plet of E$_6$ consists of ({\bf 3},{\bf 3},1$) + (${\bf 3}$^*$,1,{\bf
3}$^*) + ($1,{\bf 3}$^*$,{\bf 3}) of SU(3)$_C \times$ SU(3)$_L \times$
SU(3)$_R$, i.e., a color triplet of quarks, a color antitriplet of antiquarks,
and a nonet of color-singlet leptons. The electromagnetic charge $Q$ is then
given by 
\beq \label{eqn:qem}
Q = I_{3L} + \frac{Y_W}{2} \equiv I_{3L} + I_{3R} + \frac{Y_L + Y_R}{2}~~~.
\eeq
Unnormalized charges corresponding to U(1)$_\chi$ and U(1)$_\psi$ may be
expressed \cite{E6SS} as
\beq \label{eqn:chipsi}
Q_\chi = 4 I_{3R} - 3(Y_L + Y_R)~~,~~~Q_\psi = 3(Y_R - Y_L)~~~,
\eeq
while a charge corresponding to U(1)$_\eta$ is a linear combination of these
\cite{E6SS}: 
\beq
Q_\eta = 3 I_{3R} - 6 Y_L + (3/2) Y_R~~~.
\eeq

The authors of Ref.~\cite{BKM} note that it is possible to include in the
Lagrangian a term mixing the field strength $B_{\mu \nu}$ of weak hypercharge
U(1)$_{Y_W}$ with the field strength $X_{\mu \nu}$ of another abelian group
U(1)$_X$ without violating either U(1) symmetry.  This term can arise in
higher order of perturbation theory as a result of mixing induced by
loops of fermions with non-degenerate masses.  Thus, it is permissible to
take any linear combination of $Q_\chi$ and $Q_\psi$ and add to it a term
proportional to $Y_W = 2I_{3R} + Y_L + Y_R$ in order to try to cancel out
couplings to conventional leptons.  By this means one can construct a $Z'$
that is particularly elusive in direct searches but whose effects can be
manifested in other ways \cite{Rbb,pT}.

The assignments of quantum numbers to left-handed members of the {\bf 27}-plet
of E$_6$ are shown in Table 1.  The (unnormalized) charge $Q'$ is defined as
that linear combination of $I_{3R}$, $Y_L$, and $Y_R$ for which $Q'(e_L^-)
= Q'(\nu_{eL}) = Q'(e^+) = 0$.  Adopting a convenient normalization, we find
\beq \label{eqn:qp}
Q' = (Q_\eta + Y_W)/5 = I_{3R} - Y_L + (1/2)Y_R~~~.
\eeq
Values of this charge are also shown in Table 1.  The decoupling from leptons
of the linear combination (\ref{eqn:qp}) was noted in Refs.~\cite{delA}. 

\begin{table}
\caption{Assignment of quantum numbers to left-handed members of the {\bf
27}-plet of E$_6$.} 
\begin{center}
\begin{tabular}{c r c r r r r r r} \hline \hline
(SO(10), SU(5)) & $Q_\eta$ &
    State    & $Q$ & $I_{3L}$ & $I_{3R}$ & $Y_L$ &  $Y_R$ & $Q'$ \\ \hline
({\bf 16}, {\bf 5}$^*$) &  1   &
     $d^c$   &   1/3  &   0~   &   1/2  &   0~   & $-1/3$ &  1/3  \\
 & & $e^-$   &  $-1~$ & $-1/2$ &   0~   & $-1/3$ & $-2/3$ &  0~   \\
 & & $\nu_e$ &   0~   &   1/2  &   0~   & $-1/3$ & $-2/3$ &  0~   \\ \hline
({\bf 16}, {\bf 10}) & $-2$ &
      $u$    &   2/3  &   1/2  &   0~   &   1/3  &   0~   & $-1/3$ \\
 & &  $d$    & $-1/3$ &   1/2  &   0~   &   1/3  &   0~   & $-1/3$ \\
 & & $u^c$   & $-2/3$ &   0~   & $-1/2$ &   0~   & $-1/3$ & $-2/3$ \\
 & & $e^+$   &   1~   &   0~   &   1/2  &   2/3  &   1/3  &   0~   \\ \hline
({\bf 16}, {\bf 1}) & $-5$ &
     $N_e^c$ &   0~   &   0~   & $-1/2$ &   2/3  &   1/3  & $-1~$ \\ \hline
({\bf 10}, {\bf 5}$^*$) &  1  &
     $h^c$   &   1/3  &   0~   &   0~   &   0~   &   2/3  &  1/3  \\
 & & $E^-$   &  $-1~$ & $-1/2$ & $-1/2$ & $-1/3$ &   1/3  &  0~   \\
 & & $\nu_E$ &   0~   &   1/2  & $-1/2$ & $-1/3$ &   1/3  &  0~   \\ \hline
({\bf 10}, {\bf 5}) &  4  &
      $h$    & $-1/3$ &   0~   &   0~   & $-2/3$ &   0~   &   2/3  \\
 & & $E^+$   &   1~   &   1/2  &   1/2  & $-1/3$ &   1/3  &   1~  \\
 & &$\nu_E^c$&   0~   & $-1/2$ &   1/2  & $-1/3$ &   1/3  &   1~  \\ \hline
({\bf 1}, {\bf 1}) & $-5$ &
      $n$    &   0~   &   0~   &   0~   &   2/3  & $-2/3$ & $-1~$ \\
\hline \hline
\end{tabular}
\end{center}
\end{table}

It is amusing that the charges $Q'$ are just a re-arranged version of the
electromagnetic charges in the {\bf 27}-plet.  One passes from $Q$ in
Eq.~(\ref{eqn:qem}) to $Q'$ in Eq.~(\ref{eqn:qp}) by the substitution $I_{3L} +
(1/2)Y_L \to -Y_L$, which amounts to a Weyl reflection interchanging the first
($u$) and third ($h$) components of SU(3)$_L$. 

The values of $Q'$ in Table 1 vanish for the left-handed exotic lepton $E^-$
and its left-handed neutrino state $\nu_E$ as well as for the conventional
leptons.  However, they are largest in magnitude for all the other exotic
leptons:  the ``right-handed neutrino'' whose left-handed state is $N_e^c$, the
states $E^+$ and $\nu_E^c$, and the otherwise elusive $n$ (whose charge and
weak hypercharge both vanish, so it doesn't couple to the photon {\it or} the
standard $Z$). 

A complete set of fermions in the {\bf 27} must remain light in order to cancel
the anomaly in the charge $Q'$ \cite{BKM}.  Thus, it makes sense to imagine
that a $Z'$ coupling to this charge will have branching ratios given by
comparing the square of each charge in Table 1 to the sum of their squares.
Summing over left-handed particles and their charge-conjugates, and taking
account of color factors for quarks, we obtain the results in Table 2.  Only
single entries are shown in the second column for the Majorana particles
$N_e^c$ and $n$. If three full {\bf 27}-plets are sufficiently light, the
branching ratios in Table 2 should be divided by 3 to get each net branching
ratio (shown in the last column).  All branching ratios are reduced further if
one must take account of decays to light superpartners \cite{sp}. 

Also shown in Table 2 are forward-backward asymmetries for the quark
subprocesses $u \bar u \to f \bar f$ at the $Z'$ pole.  (Since $d$ quarks have
the same magnitude of left- and right-handed $Q'$ charges, all forward-backward
asymmetries for $d \bar d \to f \bar f$ vanish at the $Z'$ pole.)  These
asymmetries may be expressed as
\beq \label{eqn:asy}
A_{FB} = \frac{3}{4}~ \frac{[Q(u)^2 - Q(u^c)^2][Q(f)^2 - Q(f^c)^2]}
{[Q(u)^2 + Q(u^c)^2][Q(f)^2 + Q(f^c)^2]}~~~
\eeq
We have adopted the conventions that $N_e$, $h$, $E^-$, $\nu_E$, and $n$
correspond to fermions $f$.  Table 2 has a few interesting features.

\begin{table}
\caption{Branching ratios for a $Z'$ coupling to the charge $Q'$ into
various members of a single family in the {\bf 27}-plet of E$_6$.}
\begin{center}
\begin{tabular}{c c c c r} \hline \hline
State  &   Squared  & Branching & Branching & $A_{FB}(u \bar u \to$   \\
 $f$   &   charge   &   ratio   & ratio/3 (\%) & $Z' \to f \bar f)$ \\
\hline
  $d$  & $(1 + 1)/3$ &    1/12   & 2.8 & 0~~  \\
  $u$  & $(1 + 4)/3$ &    5/24   & 6.9 & 0.27 \\
$N_e^c$ &     1      &    1/8    & 4.2 & 0.45 \\
  $h$  & $(4 + 1)/3$ &    5/24   & 6.9 & $-0.27$ \\
  $E$  &   $0 + 1$   &    1/8    & 4.2 & 0.45 \\
$\nu_E$ &  $0 + 1$   &    1/8    & 4.2 & 0.45 \\
  $n$  &      1      &    1/8    & 4.2 & $-0.45$ \\ \hline
Total  &      8      &     1     & 33.3 & \\ \hline \hline
\end{tabular}
\end{center}
\end{table}

(1) In contrast to the decays of a standard $Z$, for which the branching ratio
to $d \bar d$ exceeds that to $u \bar u$, the $Z'$ considered here prefers to
decay to $u \bar u$ by a factor of 2.5.  If such a $Z'$ is heavier than $2m_t$,
it can be an additional source of top quark pairs beyond standard QCD.  A
momentum-weighted jet charge analysis \cite{mwjc} would be able to determine
whether jets produced at high transverse momenta could be due to $Z'$ decays in
which up-type species predominated. 

(2) The decays to $h$ (an exotic isosinglet quark with charge $-1/3$) are quite
prominent.  If this quark decays via flavor-changing neutral currents to other
charge $-1/3$ quarks, a signal of $Z'$ production might include unusual events
containing ordinary down-type quarks (such as $b$ quarks), photons, and virtual
or real $Z$'s. 

(3) The decays to the exotic leptons $N_e^c$, $E$, $\nu_E$, and $n$ make up
half of all $Z'$ decays to a given family.  One should then expect to see
unusual decay products consisting of leptons, photons, and virtual $Z$'s if
flavor-changing neutral currents dominate the decays of the exotic leptons.  In
principle, by a several-step mode whose details would be dependent on the
symmetry-breaking scheme giving rise to masses, a process such as $Z' \to E^+
E^-$ or $Z' \to \nu_E \bar \nu_E$ could give rise to the unusual event $\bar p
+ p \to e^+ e^- \gamma \gamma + ({\rm missing~transverse~energy})$ seen by CDF
\cite{eegg}.

(4) The prominence of up-type quark couplings to $Z'$ and the presence of
substantial forward-backward asymmetries in $u \bar u \to f \bar f$ imply
that the process $\bar p p \to Z' \to f \bar f$ is likely to produce all the
states $f$ in Table 2 except standard down-type quarks with substantial
forward-backward asymmetries.  Such asymmetries could be an early signal that
new physics is appearing through the intervention of a chiral interaction
rather than through QCD, which is left-right symmetric.

Typical searches for new $Z'$ states produced and decaying like standard $Z$'s
have reached mass limits of about 650 GeV/$c^2$ when one combines the CDF $e^+
e^-$ and $\mu^+ \mu^-$ data in samples of about 70 pb$^{-1}$ \cite{CDFZp}.  The
full sample from CDF, and the inclusion of D0 results, can be expected to more
than double the amount of data available, leading to lower limits closer to 700
GeV/$c^2$.  For $Z'$'s coupling only to U(1) factors, for which the square of
the coupling is about half of that for electroweak SU(2), one should reduce the
expected production cross sections by about a factor of 2, bringing the
anticipated limits back down to 650 GeV/$c^2$ for final states identified with
the same efficiency and branching ratio (3.4\%) as in $Z \to e^+ e^-$ decays. 
The $Z'$ discussed here has branching ratios to each species of exotic leptons
in excess of this figure, but detection efficiencies are hard to anticipate
without predictions for specific decay chains.  Indeed, to some extent it is
misleading even to identify the exotic states in E$_6$ as quarks and leptons
before we know what selection rules govern their decays. The answer to such
questions depends on symmetry-breaking schemes which we have not yet explored. 

[Note added:  After this work was completed we became aware of Ref.~\cite{GGZ},
which proposes searching for $Z' \to (W^\pm~{\rm or}~Z) + {\rm scalar}$.  In
our notation, the rates for these decays involve factors $[Q'(q_L) +
Q'(u_L^c)][Q'(q_L) + Q'(d_L^c)]$, where $q_L \equiv (u_L,d_L)$, which vanish
for our choice of charges (possibly disfavored if one seeks a solution to the
$R_b$ problem compatible with other phenomenology \cite{AGHS}).]

\section*{Acknowledgments} 

I thank the Aspen Center for Physics and the CERN and Fermilab Theory Groups
for their hospitality during parts of this study; H. Frisch, A. K. Grant, M. L.
Mangano, and J. Steinberger for fruitful discussions; and A. Faraggi, P.
Frampton, and J. L. Lopez for calling my attention to Refs.~\cite{FM},
\cite{FWW}, and \cite{LN}.  This work was supported in part by the United
States Department of Energy under Contract No.~DE FG02 90ER40560. 
 
% Journal and other miscellaneous abbreviations for references
% Phys. Lett. B style
\def \ajp#1#2#3{Am.~J.~Phys.~{\bf#1} (#3) #2}
\def \apas#1#2#3{Acta Phys.~Austriaca Suppl.~{\bf#1} (#3) #2}
\def \apny#1#2#3{Ann.~Phys.~(N.Y.) {\bf#1} (#3) #2}
\def \app#1#2#3{Acta Phys.~Polonica {\bf#1} (#3) #2}
\def \arnps#1#2#3{Ann.~Rev.~Nucl.~Part.~Sci.~{\bf#1} (#3) #2}
\def \cmp#1#2#3{Commun.~Math.~Phys.~{\bf#1} (#3) #2}
\def \cmts#1#2#3{Comments on Nucl.~Part.~Phys.~{\bf#1} (#3) #2}
\def \cn{Collaboration}
\def \corn93{{\it Lepton and Photon Interactions:  XVI International Symposium,
Ithaca, NY August 1993}, AIP Conference Proceedings No.~302, ed.~by P. Drell
and D. Rubin (AIP, New York, 1994)}
\def \cp89{{\it CP Violation,} edited by C. Jarlskog (World Scientific,
Singapore, 1989)}
\def \dpff{{\it The Fermilab Meeting -- DPF 92} (7th Meeting of the American
Physical Society Division of Particles and Fields), 10--14 November 1992,
ed. by C. H. Albright \ite~(World Scientific, Singapore, 1993)}
\def \dpf94{DPF 94 Meeting, Albuquerque, NM, Aug.~2--6, 1994}
\def \efi{Enrico Fermi Institute Report No. EFI}
\def \el#1#2#3{Europhys.~Lett.~{\bf#1} (#3) #2}
\def \f79{{\it Proceedings of the 1979 International Symposium on Lepton and
Photon Interactions at High Energies,} Fermilab, August 23-29, 1979, ed.~by
T. B. W. Kirk and H. D. I. Abarbanel (Fermi National Accelerator Laboratory,
Batavia, IL, 1979}
\def \hb87{{\it Proceeding of the 1987 International Symposium on Lepton and
Photon Interactions at High Energies,} Hamburg, 1987, ed.~by W. Bartel
and R. R\"uckl (Nucl. Phys. B, Proc. Suppl., vol. 3) (North-Holland,
Amsterdam, 1988)}
\def \ib{{\it ibid.}~}
\def \ibj#1#2#3{~{\bf#1} (#3) #2}
\def \ichep72{{\it Proceedings of the XVI International Conference on High
Energy Physics}, Chicago and Batavia, Illinois, Sept. 6--13, 1972,
edited by J. D. Jackson, A. Roberts, and R. Donaldson (Fermilab, Batavia,
IL, 1972)}
\def \ijmpa#1#2#3{Int.~J.~Mod.~Phys.~A {\bf#1} (#3) #2}
\def \ite{{\it et al.}}
\def \jmp#1#2#3{J.~Math.~Phys.~{\bf#1} (#3) #2}
\def \jpg#1#2#3{J.~Phys.~G {\bf#1} (#3) #2}
\def \lkl87{{\it Selected Topics in Electroweak Interactions} (Proceedings of
the Second Lake Louise Institute on New Frontiers in Particle Physics, 15--21
February, 1987), edited by J. M. Cameron \ite~(World Scientific, Singapore,
1987)}
\def \ky85{{\it Proceedings of the International Symposium on Lepton and
Photon Interactions at High Energy,} Kyoto, Aug.~19-24, 1985, edited by M.
Konuma and K. Takahashi (Kyoto Univ., Kyoto, 1985)}
\def \mpla#1#2#3{Mod.~Phys.~Lett.~A {\bf#1} (#3) #2}
\def \nc#1#2#3{Nuovo Cim.~{\bf#1} (#3) #2}
\def \np#1#2#3{Nucl.~Phys.~{\bf#1} (#3) #2}
\def \pisma#1#2#3#4{Pis'ma Zh.~Eksp.~Teor.~Fiz.~{\bf#1} (#3) #2 [JETP Lett.
{\bf#1} (#3) #4]}
\def \pl#1#2#3{Phys.~Lett.~{\bf#1} (#3) #2}
\def \plb#1#2#3{Phys.~Lett.~B {\bf#1} (#3) #2}
\def \pr#1#2#3{Phys.~Rev.~{\bf#1} (#3) #2}
\def \pra#1#2#3{Phys.~Rev.~A {\bf#1} (#3) #2}
\def \prd#1#2#3{Phys.~Rev.~D {\bf#1} (#3) #2}
\def \prl#1#2#3{Phys.~Rev.~Lett.~{\bf#1} (#3) #2}
\def \prp#1#2#3{Phys.~Rep.~{\bf#1} (#3) #2}
\def \ptp#1#2#3{Prog.~Theor.~Phys.~{\bf#1} (#3) #2}
\def \rmp#1#2#3{Rev.~Mod.~Phys.~{\bf#1} (#3) #2}
\def \rp#1{~~~~~\ldots\ldots{\rm rp~}{#1}~~~~~}
\def \si90{25th International Conference on High Energy Physics, Singapore,
Aug. 2-8, 1990}
\def \slc87{{\it Proceedings of the Salt Lake City Meeting} (Division of
Particles and Fields, American Physical Society, Salt Lake City, Utah, 1987),
ed.~by C. DeTar and J. S. Ball (World Scientific, Singapore, 1987)}
\def \slac89{{\it Proceedings of the XIVth International Symposium on
Lepton and Photon Interactions,} Stanford, California, 1989, edited by M.
Riordan (World Scientific, Singapore, 1990)}
\def \smass82{{\it Proceedings of the 1982 DPF Summer Study on Elementary
Particle Physics and Future Facilities}, Snowmass, Colorado, edited by R.
Donaldson, R. Gustafson, and F. Paige (World Scientific, Singapore, 1982)}
\def \smass90{{\it Research Directions for the Decade} (Proceedings of the
1990 Summer Study on High Energy Physics, June 25 -- July 13, Snowmass,
Colorado), edited by E. L. Berger (World Scientific, Singapore, 1992)}
\def \smassb{{\it Proceedings of the Workshop in $B$ Physics at Hadron
Colliders} (Snowmass, CO, June 21 -- July 2, 1993), edited by P. McBride
and C. S. Mishra, Fermilab report Fermilab-CONF-93/267 (1993)}
\def \stone{{\it B Decays}, edited by S. Stone (World Scientific, Singapore,
1994)}
\def \tasi90{{\it Testing the Standard Model} (Proceedings of the 1990
Theoretical Advanced Study Institute in Elementary Particle Physics, Boulder,
Colorado, 3--27 June, 1990), edited by M. Cveti\v{c} and P. Langacker
(World Scientific, Singapore, 1991)}
\def \yaf#1#2#3#4{Yad.~Fiz.~{\bf#1} (#3) #2 [Sov.~J.~Nucl.~Phys.~{\bf #1} (#3)
#4]}
\def \zhetf#1#2#3#4#5#6{Zh.~Eksp.~Teor.~Fiz.~{\bf #1} (#3) #2 [Sov.~Phys. -
JETP {\bf #4} (#6) #5]}
\def \zpc#1#2#3{Zeit.~Phys.~C {\bf#1} (#3) #2}


\begin{thebibliography}{99}

% 1
\bibitem{delA} F. del Aguila, G. Blair, M. Daniel, and G. G. Ross,
\np{B283}{50}{1987}; F. del Aguila, M. Qu\'{\i}ros, and F. Zwirner,
\np{B284}{530}{1987}; \ibj{287}{419}{1987}.

% 2
\bibitem{CDFZp} CDF \cn, F. Abe \ite, \prd{51}{949}{1995}; T. Kamon, Fermilab
report FERMILAB-CONF-96-106-E, hep-ex/9605006, presented at XXXI Rencontre de
Moriond:  QCD and High-energy Hadronic Interactions, 23 -- 30 March, 1996. 

% 3
\bibitem{Alt} G. Altarelli, N. di Bartolomeo, F. Feruglio, R. Gatto, and
M. L. Mangano, CERN report CERN-TH/96-29, hep-ph/9601324.

% 4
\bibitem{Chi} P. Chiappetta, J. Layssac, F. M. Renard, and C. Verzegnassi,
Montpelier Univ.~report PM-96-05, hep-ph/9601306.

% 5
\bibitem{Bam} P. Bamert, C. P. Burgess, J. M. Cline, D. London, and E. Nardi,
McGill Univ.~report McGill 96-04, hep-ph/9602438.

% 6
\bibitem{BKM} K. S. Babu, C. Kolda, and J. March-Russell, Institute for
Advanced Study report IASSNS-HEP-96/20, hep-ph/9603212; C. Kolda, 
Institute for Advanced Study report IASSNS-HEP-96/65, hep-ph/9606396,
talk given at the IV International Conference on Supersymmetry (SUSY-96),
College Park, MD, May 29 -- June 1, 1996.

% 7
\bibitem{AGHS} K. Agashe, M. Graesser, I. Hinchliffe, and M. Suzuki,
Lawrence Berkeley National Laboratory report LBL-38569, hep-ph/9604266.

% 8
\bibitem{BCL} V. Barger, K. Cheung, and P. Langacker, Univ.~of Wisconsin
report MADPH-96-936, hep-ph/9604298.

% 9
\bibitem{DTP} T. Gehrmann and W. J. Stirling, Durham Univ.~report DTP/96/24,
hep-ph/9603380; M. Heyssler, Durham Univ.~report DTP/96/42, hep-ph/9605403. 

% 10
\bibitem{FM} A. E. Faraggi and M. Masip, Univ.~of Florida report
UFIFT-HEP-96-11, hep-ph/9604302.

% 11
\bibitem{FWW} P. H. Frampton, M. B. Wise, and B. D. Wright, Univ.~of North
Carolina report IFP-722-UNC, hep-ph/9604260.

% 12
\bibitem{LN} J. L. Lopez and D. V. Nanopoulos, Rice Univ.~report
DOE/ER/40717-27, hep-ph/9605359; J. L. Lopez, Rice Univ.~report
DOE/ER/40717-30, hep-ph/9607231, to appear in Proceedings of the Fourth
International Conference on Supersymmetry (SUSY-96), College Park, MD, May 29
-- June 1, 1996. 

\bibitem{Rbb} ALEPH, DELPHI, L3, and OPAL \cn s and the LEP Electroweak Working
Group, CERN report CERN-PPE/95-172, November 1995, contributed to the 1995
International Europhysics Conference on High Energy Physics, Brussels, Belgium,
27 July -- 2 August 1995, and to the 17th International Symposium on
Lepton-Photon Interactions, Beijing, China, 10 -- 15 August 1995; ALEPH,
DELPHI, L3, and OPAL \cn s, CERN report CERN-PPE/96-017, February 1996,
submitted to Nucl.~Instr.~Meth. 

\bibitem{pT} CDF \cn, F. Abe \ite, Fermilab-Pub-96/020-E, January, 1996,
submitted to Phys.~Rev.~Letters.

\bibitem{GG} H. Georgi and S. L. Glashow, \prl{32}{438}{1974}.

\bibitem{SO} H. Georgi in {\it Proceedings of the 1974
Williamsburg DPF Meeting}, ed. by C. E. Carlson  (New York, AIP, 1975) p.~575;
H. Fritzsch and P. Minkowski, \apny{93}{193}{1975}.

\bibitem{E6SS} E. Witten, \np{B258}{75}{1985}; E. Cohen, J. Ellis, K. Enqvist,
and D. V. Nanopoulos, \pl{165B}{76}{1985}; J. L. Rosner, \cmts{15}{195}{1986}.

\bibitem{E6} F. G\"ursey, P. Ramond, and P. Sikivie, \pl{60B}{177}{1976}.

\bibitem{chipsi} R. W. Robinett, \prd{26}{2388}{1982}; R. W. Robinett and J. L.
Rosner, \prd{25}{3036}{1982}; \ibj{26}{2396}{1982}.

\bibitem{LRR} P. G. Langacker, R. W. Robinett, and J. L. Rosner,
\prd{30}{1470}{1984}.

\bibitem{LR} D. London and J. L. Rosner, \prd{34}{1530}{1986}.

\bibitem{lore} L. S. Durkin and P. Langacker, \pl{166B}{436}{1986};
U. Amaldi \ite, \prd{36}{1385}{1987};
F. del Aguila, M. Qu\'{\i}ros, and F. Zwirner, \np{B287}{457}{1987};
J. L. Hewett and T. G. Rizzo, \prp{183}{193}{1989};
P. Langacker and M. Luo, \prd{45}{278}{1992};
P. Langacker, in {\it Precision Tests of the Standard Model}, edited by
P. Langacker (World Scientific, Singapore, 1995), p.~883;
M. Cveti\v{c} and S. Godfrey, in {\it Electro-weak Symmetry Breaking and
Beyond the Standard Model}, edited by T. Barklow, S. Dawson, H. Haber, and
J. Siegrist (World Scientific, Singapore, 1995), and references therein;
M. Cveti\v{c} and P. Langacker, Institute for Advanced Study report
IASSNS-HEP-95/90, hep-ph/9511378 (unpublished).

\bibitem{sp} J. F. Gunion \ite, \ijmpa{2}{1199}{1987}; S. Nandi, \plb{197}
{144}{1987}; J. L. Hewett and T. G. Rizzo, Ref.~\cite{lore}.

\bibitem{mwjc} R. D. Field and R. P. Feynman, \np{B136}{1}{1978}.  For
recent applications see, e.g., SLD \cn, K. Abe \ite, \prl{74}{2890}{1995};
ALEPH \cn, D. Buskulic \ite, \plb{356}{409}{1995};  DELPHI \cn, P. Abreu \ite,
\zpc{65}{569}{1995}; OPAL \cn, R. Akers \ite, \zpc{67}{365}{1995}.

\bibitem{eegg} CDF \cn, F. Abe \ite, presented by S. Park, in {\it Proceedings
of the 10th Topical Workshop on Proton-Antiproton Collider Physics,} Fermilab,
May 9--13, 1995, AIP Conference Proceedings 357, edited by R. Raja and J. Yoh,
(AIP, Woodbury, NY, 1996), p.~62.

\bibitem{GGZ} H. Georgi and S. L. Glashow, Harvard University report
HUTP-96/A024, hep-ph/9607202.

\end{thebibliography}
\end{document}